\documentstyle[sprocl]{article}

\input{psfig}

\def\be{\begin{equation}}
\def\ee{\end{equation}}
\def\bea{\begin{eqnarray}}
\def\eea{\end{eqnarray}}
\begin{document}

\title{COLOR SUPERCONDUCTIVITY IN COLD, DENSE QUARK MATTER}

\author{D. H. RISCHKE}

\address{RIKEN-BNL Research Center, Brookhaven National Laboratory,\\
Upton, NY 11973, U.S.A. \\
E-mail: rischke@bnl.gov}

\author{R. D. PISARSKI} 

\address{Physics Department, Brookhaven National Laboratory,\\
Upton, NY 11973, U.S.A. \\
E-mail: pisarski@bnl.gov }

\maketitle\abstracts{We review what is different and what is similar
in a color superconductor as compared to an ordinary BCS
superconductor. The parametric dependence of the 
zero-temperature gap, $\phi_0$, on the coupling constant differs
in QCD from that in BCS theory. 
On the other hand, the transition temperature to the
superconducting phase, $T_c$, is related to the zero-temperature
gap in the same way in QCD as in BCS theory, $T_c /\phi_0 \simeq 0.567$.}

\section{Cold, dense fermionic matter}

Consider a degenerate, non-interacting fermionic system, where
all momentum states up to the Fermi surface are occupied. The 
fermion occupation number is a step function,
$ n_0(\epsilon_0) = \theta(-\epsilon_0)$,
where $\epsilon_0 \equiv \omega - \mu$ is the energy of the fermions
with respect to the Fermi energy, $\mu$,
and $\omega$ is their kinetic energy. For non-relativistic particles with
3-momentum ${\bf k}$ and mass $m$,
$\omega = {\bf k}^2/(2m)$, while for massless (ultrarelativistic) particles,
$\omega = k \equiv |{\bf k}|$. (The units are $\hbar = k_B = c =1$.)
The excitation spectrum of the fermions consists of a
particle branch, $\epsilon_0^{\rm p} \equiv \epsilon_0$, and a hole
branch, $\epsilon_0^{\rm h} \equiv - \epsilon_0$. 
The formation of particle-hole excitations at 
the Fermi surface costs no energy, $\epsilon_0^{\rm p} + \epsilon_0^{\rm h}
\equiv 0$.

Now switch on an interaction, for the sake of simplicity a
point-like four-fermion interaction with interaction strength $G^2$.
Let the sign of $G^2$ be defined such that $G^2 >0$ in the case of an 
attractive interaction, and $G^2 <0$, if the interaction is
repulsive. Due to Pauli's exclusion principle
and energy conservation, scattering of fermions can
occur exclusively {\em at\/} the Fermi surface. In other words,
physical scattering processes are possible only for $\epsilon_0
\rightarrow 0$. The amplitude for
fermion-fermion scattering is \cite{BCS}
\begin{equation} \label{scatt}
\Gamma(\epsilon_0) \sim \frac{G^2}{1- G^2 \, \ln(\mu/\epsilon_0)} \;\; ,
\end{equation}
where we suppress all constant factors that are irrelevant for the qualitative
arguments presented here.
As $\epsilon_0 \rightarrow 0$, in the
case of an attractive interaction the scattering amplitude
develops a singularity at an energy scale
\begin{equation} \label{sing}
\epsilon_0^* \sim \mu\, \exp(-1/G^2) \;\; . 
\end{equation}
For repulsive interactions, no such singularity occurs.
Obviously, the singularity (\ref{sing}) occurs even when the attractive
interaction is arbitrarily weak, $G^2 \rightarrow 0^+$,
all that changes is the energy scale $\epsilon_0^*$ of the singularity.

This singularity is the famous Cooper instability \cite{BCS}.
It is cured by the formation of Cooper pairs which, as
bosons, condense in the true, energetically favored ground state of
the system. Macroscopically, this leads to the phenomenon of
{\em superfluidity}, or, if the fermions carry charge,
{\em superconductivity}.

The excitation spectrum of the system also changes. A gap $\phi_0$
forms at the Fermi surface, separating the branch for
quasiparticle excitations, 
$\epsilon^{\rm p} = - \epsilon$, 
$\epsilon \equiv \sqrt{\epsilon_0^2 + \phi_0^2}$,
from the one for quasihole excitations,
$\epsilon^{\rm h} \equiv + \epsilon$.
Now exciting a quasiparticle--quasihole pair costs at least
an energy $\epsilon^{\rm p} + \epsilon^{\rm h} \geq 2\, \phi_0$.
The quasiparticle occupation number, $n(\epsilon_0)$, 
is ``smeared'' around the (original) Fermi surface, 
$n(\epsilon_0) = (\epsilon - \epsilon_0)/(2 \epsilon)$.

To compute the gap, one has to solve a gap
equation \cite{BCS}. In the case of a point-like four-fermion interaction,
this equation takes the form
\begin{equation} \label{gapequation}
\phi_0 = G^2 \int_{-\mu}^0 \frac{{\rm d}\epsilon_0}{\epsilon} \, \phi_0 \;\; .
\end{equation}
(Again, irrelevant constant factors are omitted.) Besides the trivial
(energetically disfavored) solution $\phi_0 = 0$,
this equation has always a non-trivial (energetically favored)
solution, $\phi_0 \neq 0$.
Since the gap is a constant for point-like four-fermion interactions,
we can divide both sides of Eq.\ (\ref{gapequation}) by $\phi_0$. 
The remaining integral can be solved exactly, with the result
\begin{equation} \label{solution}
\phi_0 \simeq 2\, \mu \, \exp (-1/G^2) \;\; .
\end{equation} 
Apparently, $\phi_0 > \epsilon_0^*$, cf.\ Eq.\ (\ref{sing}). In other words,
since there are no quasiparticle states with 
energy $\epsilon^{\rm p} \geq - \phi_0$ (or quasihole states with
$\epsilon^{\rm h} \leq \phi_0$),
the gap has just the right order of magnitude to prevent 
the scattering amplitude (\ref{scatt}) from developing a singularity.

The assumption of a point-like four-fermion interaction can be relaxed.
Assume that the interaction is mediated by a scalar boson of mass $M$.
For the sake of definiteness, assume that the boson mass is
generated by many-body interactions at nonzero density, $M \sim g \mu$.
The boson-fermion coupling is denoted by $g$, the boson propagator
is $\Delta(P)=1/(M^2-P^2)$, $P \equiv P^\mu =(p_0,{\bf p})$,
$P^2 \equiv p_0^2 - {\bf p}^2$.
In this case, the gap equation (\ref{gapequation}) becomes
\cite{rdpdhrscalar}
\begin{equation} \label{gapequation2}
\phi_0(k) = g^2 \int_{-\mu}^0 \frac{{\rm d}\epsilon_0}{\epsilon} \,
\phi_0(q) \, \frac{q}{k} \, \ln \left[ \frac{M^2 + (k + q)^2}{M^2 + (k-q)^2}
\right] \;\; .
\end{equation}
Here, $q \equiv \epsilon_0 + \mu$. The frequency dependence of the
boson propagator and, consequently, that of the gap function has 
been neglected, based on the argument that in weak coupling, $g \ll 1$,
$p_0 \sim \epsilon \sim \phi_0 \sim \mu \, \exp(-1/g^2) \ll M \sim g \mu$.
The logarithm arises from the integration over the angle between the 
boson 3-momentum ${\bf p} \equiv {\bf k} - {\bf q}$ and 
the 3-momentum of the fermion in the condensate, ${\bf q}$, for
details see Ref.\ \cite{rdpdhrscalar}. 
Note that this factor enhances contributions from the region of 
momenta ${\bf q} \simeq {\bf k}$ ({\em collinear enhancement}).

In the case of a boson-mediated interaction, the gap is no 
longer constant, but a function of momentum.
Consequently, the gap equation (\ref{gapequation2}) is no longer a
simple fix-point equation for $\phi_0$, but an integral equation which 
has to be solved numerically.
However, to estimate the order of magnitude of the gap at the Fermi
surface, $k \equiv \mu$ (for massless fermions), one may make the following
approximation. First note that,
due to the factor $1/\epsilon$, the integrand peaks at the Fermi surface.
It is then sufficient
to approximate the slowly varying logarithm and the gap function
with their values for $q = k = \mu$. This leads to the estimate
\begin{equation} \label{solution2}
\phi_0 \simeq 2\, \mu\, \exp \left[ - \frac{1}{g^2\, \ln 
(1 + 2\mu^2/M^2)} \right]
\end{equation}
for the value of the gap function at the Fermi surface, $\phi_0 \equiv 
\phi_0(\mu)$,
which should be compared with Eq.\ (\ref{solution}). All that changed
is that the coupling constant is effectively increased by the logarithm
originating from collinear enhancement. For $M \sim g \mu$,
the logarithm becomes $\sim \ln(1/g)$ in weak coupling.

\section{Cold, dense quark matter}

The density in cold quark matter increases $\sim \mu^3$.
Asymptotic freedom then implies that
single-gluon exchange becomes the dominant interaction between
quarks. Single-gluon exchange is attractive in the color-antitriplet channel, 
and therefore leads to color superconductivity in cold, dense quark matter 
\cite{BarroisPhD,others,bailinlove}. 
Recently, considerable activity was generated $^{6-39}$
by the work of Refs.\ \cite{ARW,RSSV}, which suggested that
the zero-temperature color-superconducting gap $\phi_0$
could be as large as 100 MeV.
This order of magnitude is quite surprising, because earlier
work by Bailin and Love \cite{bailinlove} estimated
the gap to be $\phi_0 \sim 1$ MeV.
While gaps of order 100 MeV could also be relevant for the physics of
nuclear collisions (see below), gaps of about 1 MeV allow at most for
the possibility that neutron star cores, if consisting of quark matter, 
could be color superconductors.

The authors of Refs.\ \cite{ARW,RSSV} based their arguments on a simple model
where quarks interact via a point-like four-fermion interaction, 
giving rise to a gap equation of the type (\ref{gapequation}).
The coupling strength $G^2$ was adjusted such that the model reproduced
the order of magnitude of the chiral transition at nonzero temperature
and zero quark chemical potential, $T_\chi \sim 150$ MeV.
The earlier work of Bailin and Love \cite{bailinlove} was already more
sophisticated in the sense that they used
one-gluon exchange, employing gluon propagators with electric and
magnetic screening masses.
This gives rise to a gap equation of
the form (\ref{gapequation2}), with $g$ being the strong coupling constant.

Unfortunately, both approaches fail to capture an essential property of
single-gluon exchange: at zero temperature, due to the absence of
magnetic screening, magnetic interactions are
truly long-range. Surprisingly, this fact was already known to
Barrois \cite{BarroisPhD}, but apparently never made it into
the published literature.

Long-range magnetic interactions have the important consequence that
one can no longer neglect the frequency dependence of the boson propagator,
as done in the derivation of (\ref{gapequation2}). For massless boson
exchange, the gap equation assumes the (approximate) form
\begin{equation} \label{gapequation3}
\phi_0(k_0) \simeq g^2 \int_{-\mu}^0 \frac{{\rm d}\epsilon_0}{\epsilon} \,
\phi_0(\epsilon)\, \frac{1}{2}
\, \ln \left( \frac{ \mu^2}{\epsilon^2-k_0^2} \right)
\;\; .
\end{equation}
Again, as in (\ref{gapequation2}) there is an additional logarithm,
representing collinear enhancement. In QCD it arises from the
exchange of ultrasoft, magnetic gluons. 
While the collinear enhancement in Eq.\ (\ref{gapequation2}) is cut off
by the mass of the scalar boson, $M$, here it is cut off
by the {\em energy\/} of the magnetic gluon, $p_0 = \epsilon \pm k_0$.
For $\epsilon \sim k_0 \sim \phi_0$, $p_0$ is of the order of $\phi_0$, too,
while in weak coupling, $M \sim g \mu \gg \phi_0$. 
For magnetic gluon exchange in QCD, the contribution of the 
collinear region, ${\bf q} \simeq {\bf k}$, to the gap integral
is therefore much larger than in the case of massive boson exchange.

To estimate the effect of the logarithm on the
solution of the gap equation,
let us neglect the energy dependence of the gap function
in the integrand and consider its value at the Fermi surface,
$k = \mu$, $k_0 = \phi_0$. One may also make the approximation \cite{son}
$\ln [\mu^2/(\epsilon^2 - \phi_0^2)] \simeq 2\, \ln(\mu/\epsilon)$.
Then, the integral is again
exactly solvable, with the (order of magnitude) result
\begin{equation} \label{solution3}
\phi_0 \simeq 2\,  \mu \, \exp(-1/g)\;\; .
\end{equation}
Due to the explicit $\epsilon$ dependence of the logarithm in
(\ref{gapequation3}), the
power of $g$ in the exponent is reduced as compared to the BCS result
(\ref{solution}). The case of
massive scalar boson exchange, Eq.\ (\ref{solution2}),
interpolates between these two
cases, as $g^2 \ll g^2 \ln(1/g) \ll g$ for $g \ll 1$.
This reflects the fact mentioned earlier that, while
the collinear singularity ${\bf q} = {\bf k}$ 
in (\ref{gapequation2}) is cut off by the mass of the scalar boson, 
$M \sim g \mu$, the singularity in (\ref{gapequation3}) is
cut off by the gluon energy, $p_0 \sim \phi_0$, which is,
in weak coupling, much smaller than $M$.
In the literature, the parametric dependence on $g$ of the solution
(\ref{solution3}) to the gap equation (\ref{gapequation3}) was
first discussed in Refs.\ \cite{rdpdhrlett1,hong,son}.

As of today, various refinements of
the solution (\ref{solution3}) have been
discussed \cite{SWQCD,rdpdhr2,OhioQCD,son,rockef}. 
The value of the gap function at the Fermi surface, $k = \mu$, $k_0 = \phi_0$,
is
\begin{equation} \label{refinedsol}
\phi_0 = b\, \mu\, \, g^{-5} \, \exp\left(-\frac{c}{g}\right) \,
\left[1+O(g)\right] \;\;\;\; ,
\;\;\;\; c= \frac{3 \, \pi^2}{\sqrt{2}}\;\; .
\end{equation}
Furthermore, the gap function has a non-trivial (on-shell) energy dependence,
\begin{equation} \label{energydep}
\phi_0(\epsilon) = \phi_0 \, \sin \left[ \frac{\pi}{2} \, \frac{g}{c}
\, \ln \left( \frac{b\, \mu}{f(\epsilon)} \right) \right] \;\; .
\end{equation}
The constant $c$ was first computed by Son \cite{son}.
To obtain the correct numerical value for $c$,
one has to account for the modifications of
the gluon propagator in the presence of a dense
medium \cite{LeBellac}. Then, what dominates the
gap equation (\ref{gapequation3}) and determines $c$ is
the contribution from nearly static, Landau-damped 
magnetic gluons. Furthermore, Son showed that in computing $c$,
it is essential
to retain the energy dependence (\ref{energydep}) of the gap function.
To the level of accuracy considered by Son \cite{son}, $b=1$ and $f(\epsilon)
\equiv \epsilon$. The prefactor $g^{-5}$ arises from
subleading contributions of static electric and non-static magnetic
gluons.

The constant $b$ collects constant factors in these subleading
contributions. It was first computed by Sch\"afer and Wilczek \cite{SWQCD}
and the present authors \cite{rdpdhr2},
\begin{equation} \label{b}
b = 512 \, \pi^4 \, \left( \frac{2}{N_f} \right)^{5/2} \, b' \;\; ,
\end{equation}
with an undetermined constant $b'$. 
Sch\"afer and Wilczek \cite{SWQCD} obtained $b' =1/2$ and, as 
before, $f(\epsilon) \equiv \epsilon$. 
We showed \cite{rdpdhr2} that actually $b' = 1$ 
and $f(\epsilon) \equiv \epsilon - \epsilon_0$.
The additional factor 2 is the same that occurs in Eqs.\
(\ref{solution}), (\ref{solution2}), and (\ref{solution3}).
It arises from the measure of integration in the gap equation,
${\rm d}\epsilon_0/\epsilon \equiv - {\rm d} \ln (
\epsilon - \epsilon_0)$, and $- \ln (\epsilon - \epsilon_0) |_{ -\mu}^0 
\simeq \ln (2\mu/\phi_0)$. This also explains the modification
of $f(\epsilon)$.
In addition, we pointed out \cite{rdpdhr2} that it is
important to compute the gap function on the correct quasiparticle
mass shell, and we considered for the first
time the case of non-zero temperature. As will be discussed in more detail
below, the temperature $T_c$ where the color-superconducting condensate
melts is related to the zero-temperature gap in the same way as
in BCS theory, $T_c/\phi_0 = e^\gamma/\pi \simeq 0.567$, where
$\gamma \simeq 0.577$ is the Euler--Mascheroni constant.

\begin{figure}[t]
\hspace*{0.5cm}
\psfig{figure=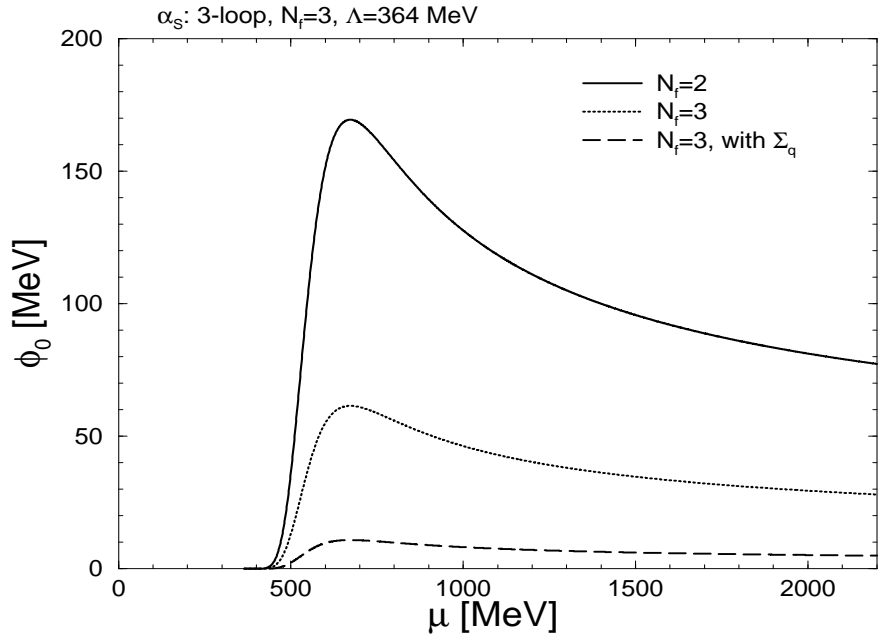,width=2.9in,height=3.5in,angle=-90}
\vspace{-1.5cm}
\caption{The value of the zero-temperature
gap at the Fermi surface, $\phi_0$,
as a function of the quark chemical potential $\mu$ for
2 quark flavors (full line), 3 quark flavors (dotted line),
and 3 quark flavors including effects from the quark wavefunction
renormalization in the gap equation (dashed line).}
\label{fig1}
\end{figure}

Brown, Liu, and Ren \cite{rockef} 
computed $T_c$ from the quark-quark scattering
amplitude. Using the result $T_c / \phi_0 = e^\gamma/\pi$
of Ref.\ \cite{rdpdhr2}, they obtain
\begin{equation}
b' = \exp\left(-\frac{\pi^2+4}{8}\right) \simeq 0.176 \;\; ,
\end{equation}
where the correction to the previous estimate $b'=1$ arises from 
a finite, $\mu$ dependent 
contribution to the wavefunction renormalization for quarks in
a dense medium. 
The authors of Ref.\ \cite{rockef} also assert that there
are no further corrections to $b'$ at this order in $g$.
If correct, this is a remarkable result, because from previous calculations
\cite{SWQCD,rdpdhr2} it appeared that computing $b'$ exactly
to leading order in $g$ would be a formidable task. 

Unlike calculations of the free energy or the Debye mass,
where the perturbative expansion in powers of $g^2$ appears to be
well-behaved \cite{rdpdhr2} 
even when extrapolated down to moderate values of $\mu$,
this result for the wavefunction renormalization (which is equivalent
to non-Fermi liquid behavior) indicates that perturbation theory is
not such a good approximation, at least for quarks near the Fermi surface.
The confirmation of the results of Ref.\ \cite{rockef} is clearly an 
outstanding problem for the field.

Figure \ref{fig1} shows the value of the zero-temperature gap 
$\phi_0$ at the Fermi surface according to Eq.\ (\ref{refinedsol})
as a function of $\mu$ for $N_f=2$ and $3$ massless quark flavors
with $b' = 1$ (full and dotted lines), and for $N_f=3$ flavors
with $b' = \exp[-(\pi^2+4)/8]$ (dashed line).
The running of the coupling $g(\mu)$ with the chemical potential
$\mu$ was computed from the 3-loop QCD $\beta$ function
\cite{PDB}, however,
not for 6 but only for 3 flavors of massless quarks. 
Therefore, the QCD scale $\Lambda = 364$ MeV is chosen 
somewhat larger than the standard value, to give the value
$\alpha_s(\mu=2\, {\rm  GeV}) \simeq 0.309$.
Although an extrapolation of the weak-coupling result
(\ref{refinedsol}) to large $g$ (small $\mu$) appears audacious, it
is interesting to note that the maximum value of the gap for
$b' = 1$ is of the order of 100 MeV, quite in agreement with the
earlier estimates of Refs.\ \cite{ARW,RSSV}. 
However, taking into account the quark
wavefunction renormalization \cite{rockef} reduces the gap to 
values of a few MeV.
These values are of the order of typical superfluid gaps in ordinary
hadronic matter. This lends credibility to
the conjecture that quark and hadronic
matter are continuously connected \cite{cont}, although 
symmetry arguments \cite{rdpdhrlett1,rob} suggest that, at zero temperature, 
there is a first order phase transition between these two phases of nuclear
matter.

\section{Not so cold, dense quark matter}

To understand how the color-superconducting gap changes with temperature,
it is instructive to first consider the simpler BCS case. 
At nonzero temperature $T$, the gap equation (\ref{gapequation})
becomes
\begin{equation} \label{gapequationT}
\phi = G^2 \int_{-\mu}^0 \frac{{\rm d}\epsilon_0}{\epsilon(\phi)} \, \phi 
\, \tanh \left[ \frac{\epsilon(\phi)}{2T} \right] \;\; .
\end{equation}
Here, $\phi$ is the value of the gap at temperature $T$,
$\phi \equiv \phi(T)$, and as before, $\phi_0 \equiv \phi(0)$ 
denotes the zero-temperature gap in the following.
The $\phi$ dependence of the quasiparticle excitation energy
$\epsilon(\phi) \equiv \sqrt{\epsilon_0^2 + \phi^2}$ 
has been made explicit, to distinguish $\epsilon(\phi)$ from
$\epsilon \equiv \epsilon(\phi_0)$ used previously.

Again, $\phi$ is constant and can be divided out on both sides
of Eq.\ (\ref{gapequationT}), with the result
\begin{equation} \label{gapequationT2}
1 = G^2 \int_{-\mu}^0 \frac{{\rm d}\epsilon_0}{\epsilon(\phi)} \, 
\, \tanh \left[ \frac{\epsilon(\phi)}{2T} \right] \;\; .
\end{equation}
The effect of the $\tanh$ is to reduce the value of the integrand,
such that $\phi$ in the factor $1/\epsilon(\phi)$
has to decrease in order to balance the 1 on the
left-hand side. At some critical temperature $T_c$, this balance can
no longer be achieved and $\phi=0$ is the only solution of Eq.\
(\ref{gapequationT}). $T_c$ is the temperature where the
superconducting condensate melts. Physically, the random thermal energy
of the fermions exceeds their binding energy in a Cooper pair.
Thus, $T_c$ must be of the same order as $\phi_0$.

There is an easy way to compute the change of $\phi$ with $T$, which was
to our knowledge first suggested in Ref.\ \cite{rdpdhr2}.
Note that $\tanh [\epsilon(\phi)/2T] \simeq 1$
far from the Fermi surface, $\epsilon(\phi) \geq |\epsilon_0| 
\gg \phi_0 \sim T$.
A nonzero temperature influences the integrand only in the region
close to the Fermi surface, $|\epsilon_0| \leq  \phi_0 \sim T$.
Let us therefore divide the range of integration into two parts,
$0 \geq \epsilon_0 \geq - \kappa\, \phi_0$, 
and $- \kappa\, \phi_0 \geq \epsilon_0 \geq - \mu$,
where $\kappa \gg 1$. Then, the $\tanh$ need only be kept in the
first region, and the integral over the second region can be performed
similarly as at zero temperature,
\begin{equation}
1 \simeq G^2 \int^0_{- \kappa\, \phi_0} 
\frac{{\rm d}\epsilon_0}{\epsilon(\phi)} \, 
\, \tanh \left[ \frac{\epsilon(\phi)}{2T} \right] + 
G^2 \, \ln \left( \frac{ \mu}{ \kappa \, \phi_0} \right)\;\; .
\end{equation}
Using the solution (\ref{solution}) for the zero-temperature
gap $\phi_0$, the second term becomes  $1-G^2\, \ln (2\, \kappa)$.
Apparently, the 1 on the left-hand side is almost completely saturated
by this second term. Cancelling the 1 and writing 
$\ln (2 \, \kappa) \equiv \int^0_{-\kappa \, \phi_0} {\rm d} \epsilon_0/
\epsilon(\phi_0)$, one obtains the condition
\begin{equation} \label{condition}
G^2 \int_{-\kappa\, \phi_0}^0 {\rm d}\epsilon_0 \left\{
\frac{1}{\epsilon(\phi)} \, 
\, \tanh \left[ \frac{\epsilon(\phi)}{2T} \right] - 
\frac{1}{\epsilon(\phi_0)} \right\} = 0 \;\; .
\end{equation}
The dependence on $\kappa$ is spurious: one might
as well send $\kappa \rightarrow \infty$~\cite{rdpdhr2}, because the integrand
vanishes when $\kappa \gg 1$.
Equation (\ref{condition}) determines $\phi(T)/\phi_0$ as a function of $T$.
In particular, at $T_c$, where $\phi=0$, one derives the well-known
result
\begin{equation} \label{ratio}
\frac{T_c}{\phi_0} = \frac{e^{\gamma}}{\pi} \simeq 0.567
\end{equation}
mentioned above.

In QCD, it turns out \cite{rdpdhr2} that the effect of temperature on the
gap equation is essentially identical to that in BCS theory:
the integrand in (\ref{gapequation3}) is multiplied with
$\tanh[\epsilon(\phi)/2T]$. 
For the same reasons as in the BCS case, this factor is
negligible far away from the Fermi surface. 
One may again divide the range of integration into two parts, and neglect the
effects of temperature in the one far from the Fermi surface.
The integral over this region can be computed as for $T=0$.
Quite similarly to the treatment in the BCS case,
it is found to saturate the left-hand side of the gap equation
up to corrections of order $O(g)$.

Therefore, the integral over the region close to the Fermi surface
must also be of order $O(g)$, in order to cancel these corrections. To see
this, it is permissible to compute this integral to leading order in $g$.
One then derives the same condition (\ref{condition}) as in the BCS case, 
except that $G^2$ is replaced by $g$,
see Ref.\ \cite{rdpdhr2} for details.
Consequently, to leading order in $g$,
the $T$ dependence of the gap at the Fermi surface, 
normalized to the zero-temperature gap,
$\phi(T)/\phi_0$, is the same as in BCS theory. In particular, the
ratio $T_c/\phi_0$ is again given by Eq.\ (\ref{ratio}).
In retrospect, this is not surprising: the prefactor $b$ of the
zero-temperature gap, Eq.\ (\ref{b}), was seen to be determined by subleading 
terms in the gap equation [terms of order $O(g)$ 
relative to the leading terms due to Landau-damped magnetic gluons]. 
As explained above, temperature affects the gap equation at the
same subleading order.

An immediate consequence is that when multiplying the ordinate of Fig.\ 1
with $0.567$, one obtains the location of the phase transition to the
color-superconducting phase in the $T-\mu$ phase diagram of nuclear matter.
In the 2-flavor case without wavefunction corrections,
the transition temperature is of order $\sim 100$ MeV. The 
color-superconducting phase could then be accessible in heavy-ion collisions
at BNL--AGS or GSI--SIS energies, which explore the range of moderate
temperatures and high (net) baryon density in the nuclear
matter phase diagram. However, in the 3-flavor case, 
including the effects of the quark wavefunction renormalization,
the transition temperature is at most $\sim 6$ MeV.
For such small temperatures, color superconductivity occurs at best in
neutron star cores, if they consist of quark matter.

\section*{Acknowledgements}
D.H.R.\ would like to thank the organizers of QCD 2000 for the invitation
to give this talk. The authors thank
T.\ Sch\"afer and D.T.\ Son for many enlightening discussions.
D.H.R.\ thanks RIKEN, BNL and the U.S.\ Dept.\ of Energy for
providing the facilities essential for the completion of this work,
and to Columbia University's Nuclear Theory Group for
continuing access to their computing facilities.

\end{document}